\documentclass{emulateapj}
\bibpunct{(}{)}{;}{a}{}{,}

\newcommand{\rh}{\varrho}
\newcommand{\vunit}{\mbox{m}\,\mbox{s}^{-1}}
\newcommand{\dunit}{\mbox{m}^2\,\mbox{s}^{-1}}

\begin{document}

\title{Origin of solar torsional oscillations}

\author{Matthias Rempel}

\affil{High Altitude Observatory,
       National Center for Atmospheric Research\footnote{The National
       Center for Atmospheric Research is sponsored by the National
       Science Foundation} , 
       P.O. Box 3000, Boulder, Colorado 80307, USA
      }

\email{rempel@hao.ucar.edu}

\shorttitle{Torsional Oscillations}
\shortauthors{M. Rempel}

\begin{abstract}
  Helioseismology has revealed many details of solar differential rotation
  and also its time variation, known as torsional oscillations. So far there
  is no generally accepted theoretical explanation for torsional oscillations,
  even though a close relation to the solar activity cycle is evident. On the
  theoretical side non-kinematic dynamo models (including the Lorentz force 
  feedback on differential rotation) have been used to explain torsional 
  oscillations. In this paper we use a slightly different approach by forcing
  torsional oscillations in a mean field differential rotation model. Our aim
  is not a fully self-consistent model but rather to point out a few general
  properties of torsional oscillations and their possible origin that are
  independent from a particular dynamo model. We find that the poleward
  propagating high latitude branch of the torsional oscillations can be 
  explained as a response of the coupled differential rotation / meridional 
  flow system to periodic forcing in mid-latitudes, of either mechanical 
  (Lorentz force) or thermal nature. The speed of the poleward propagation 
  sets constraints on the value of the turbulent viscosity in the solar 
  convection zone to be less than $3\times 10^8\,\dunit$. We also show that the 
  equatorward propagating low latitude branch is very unlikely a consequence
  of mechanical forcing (Lorentz force) alone, but rather of thermal origin
  due to the Taylor-Proudman theorem. 
\end{abstract}


\section{Introduction}
Solar torsional oscillations have been known to exist for more than two 
decades. \citet{Howard:Labonte:1980} presented the first observations of
torsional oscillations using Mt. Wilson Doppler measurements and
pointed out the 11 year periodicity and the relation to the solar
cycle. These early observations showed only the equatorward 
propagating branch at low latitudes. The high latitude branch 
(above $60\degr$), which is in amplitude at least twice as strong as the 
equatorward propagating branch, was found more recently through helioseismic 
measurements by \citet{Toomre:etal:2000}, \citet{Howe:etal:2000:solphys},
\citet{Antia:Basu:2001}, \citet{Vorontsov:etal:2002}, and 
\citet{Howe:etal:2005}. These inversions also show that the high
latitude signal penetrates almost all the way to the base of the convection 
zone. The depth penetration of the low latitude signal is more uncertain due
to the lower amplitude, which is comparable to the uncertainties of the 
inversion methods in the lower half of the convection zone.
The most interesting feature of the low latitude branch is an inclination of 
the phase with respect to the rotation axis by about $25\degr$. This 
inclination is very close to the inclinations of the isorotation contours of 
$\Omega$ as pointed out by \citet{Howe:etal:2004:phase}.  

On the theoretical side a variety of explanations have been proposed:
macroscopic Lorentz force feedback, microscopic Lorentz force feedback, and
thermal forcing.

The idea of macroscopic Lorentz force feedback (computed from the
large scale magnetic mean field of the solar dynamo) was originally
proposed by \citet{Schuessler:1981} and \citet{Yoshimura:1981} and has been 
incorporated into dynamo models more recently by 
\citet{Covas:etal:2000,Covas:etal:2004,Covas:etal:2005}. 
While these models address the non-linear Lorentz force feedback using a
simplified equations of motion (considering only the longitudinal component), 
models by \citet{Jennings:1993} and \citet{Rempel:2006:dynamo} 
consider the Lorentz force feedback also in the meridional plane. The model of 
\citet{Rempel:etal:2005}, \citet{Rempel:2006:dynamo} is along the lines of the 
$\alpha\Lambda$-models by \citet{Brandenburg:etal:1990,Brandenburg:etal:1991,
Brandenburg:etal:1992}, \citet{Moss:etal:1995}, and \citet{Muhli:etal:1995} 
(coupling mean field 
models for differential rotation, meridional flow and magnetic field 
evolution), but puts more emphasis on the role of the meridional flow 
leading to a flux-transport dynamo (see e.g. \citet{Dikpati:2005} for a recent
review on the development of flux-transport dynamos). 

Microscopic Lorentz force feedback (quenching of turbulent transport processes
driving differential rotation '$\Lambda$-quenching') has been addressed by
\citet{Kitchatinov:Pipin:1998}, \citet{Kitchatinov:etal:1999}, and
\citet{Kueker:etal:1999}.

Very recently \citet{Spruit:2003} proposed a thermal origin of the low latitude
branch of torsional oscillations, driven through enhanced radiative losses in 
the active region belt. This theory also predicts an inflow into the active
region belt, which has been observed by \citet{Komm:etal:1993}, 
\citet{Komm:1994}, and \citet{Zhao:Kosovichev:2004}. 

The investigation presented here is based on the model of 
\citet{Rempel:2006:dynamo}; however we take a different 
approach by decoupling the excitation of torsional oscillations from a detailed
dynamo model. We use the differential rotation model of \citet{Rempel:2005}
and force torsional oscillations through mechanical and thermal perturbations
to address the question of what type of forcing is required to get a response
comparable to the observed torsional oscillations. 

\begin{figure*}
  \resizebox{\hsize}{!}{\includegraphics{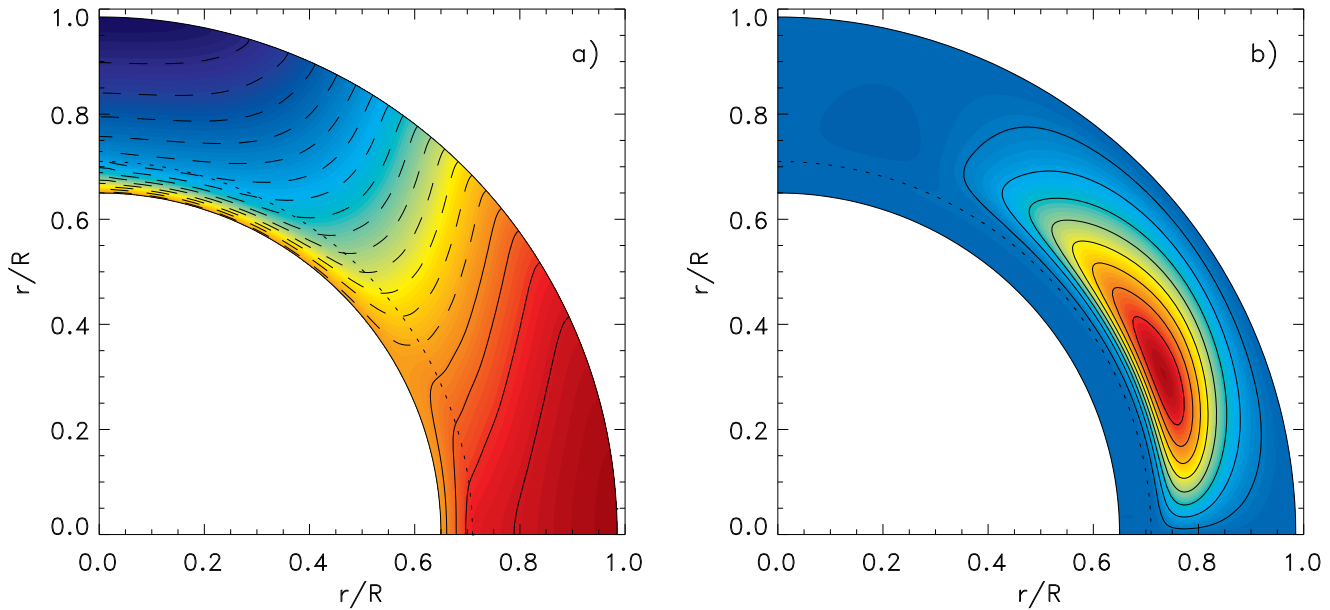}}
  \caption{a) Differential rotation (solid contours correspond to faster 
    rotation, dashed contours to slower rotation than the core of the sun); 
    b) Stream function of meridional flow (solid contours correspond to a 
    counter clockwise flow). The amplitude of differential rotation is $30\%$
    of the core rotation rate, the amplitude of the meridional flow is 
    $11\,\vunit$ ($2\,\vunit$) at the top (bottom) of the convection zone.
  }
  \label{f1}
\end{figure*}

\begin{figure*}
  \resizebox{\hsize}{!}{\includegraphics{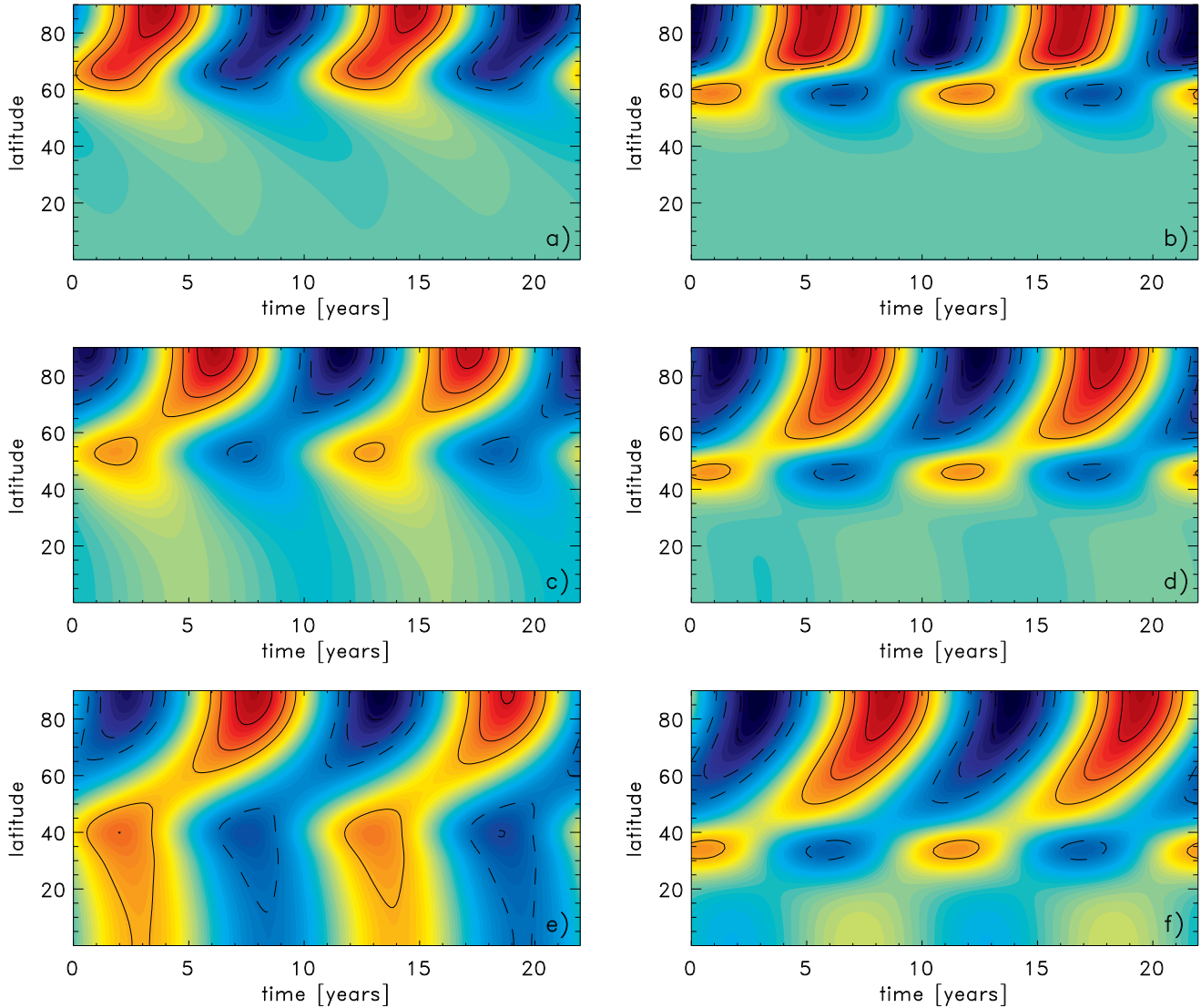}}
  \caption{Torsional oscillations produced by periodic perturbations at a 
    fixed location. The left panels (a, c, e) show the response to a periodic
    perturbation in the angular velocity (mechanical forcing), the right 
    panels (b, d, f) the response to a periodic perturbation in entropy 
    (thermal
    forcing). Closed contours correspond to positive values (faster rotation),
    dashed contours to negative values (slower rotation). From top to bottom 
    the position of the perturbation in latitude is varied between $60\degr$ 
    (a, b), $45\degr$ (c, d), and $30\degr$ (e, f). In all cases the 
    strongest signal is found at the pole, independent from the location of
    the forcing. The power on the poleward side is always higher than on the
    equatorward side. 
  }
  \label{f2}
\end{figure*}

We emphasize that this is not a fully self-consistent explanation of torsional
oscillations since we do not incorporate a full dynamo model also considering the
non-linear feedback of zonal and meridional flow variations on the magnetic field
evolution. We have published such a self-consistent non-linear dynamo model (based on a 
flux-transport dynamo and the same differential rotation model we use in this paper) 
in \citet{Rempel:2006:dynamo}. We found that many features of torsional oscillations
discussed in that paper are more general, meaning independent from a particular dynamo
model. The goal of this paper is to point out these general properties that are strongly
related to influence of the Taylor-Proudman theorem on amplitude and phase of torsional
oscillations.   

\section{Model}

\begin{figure*}
  \resizebox{\hsize}{!}{\includegraphics{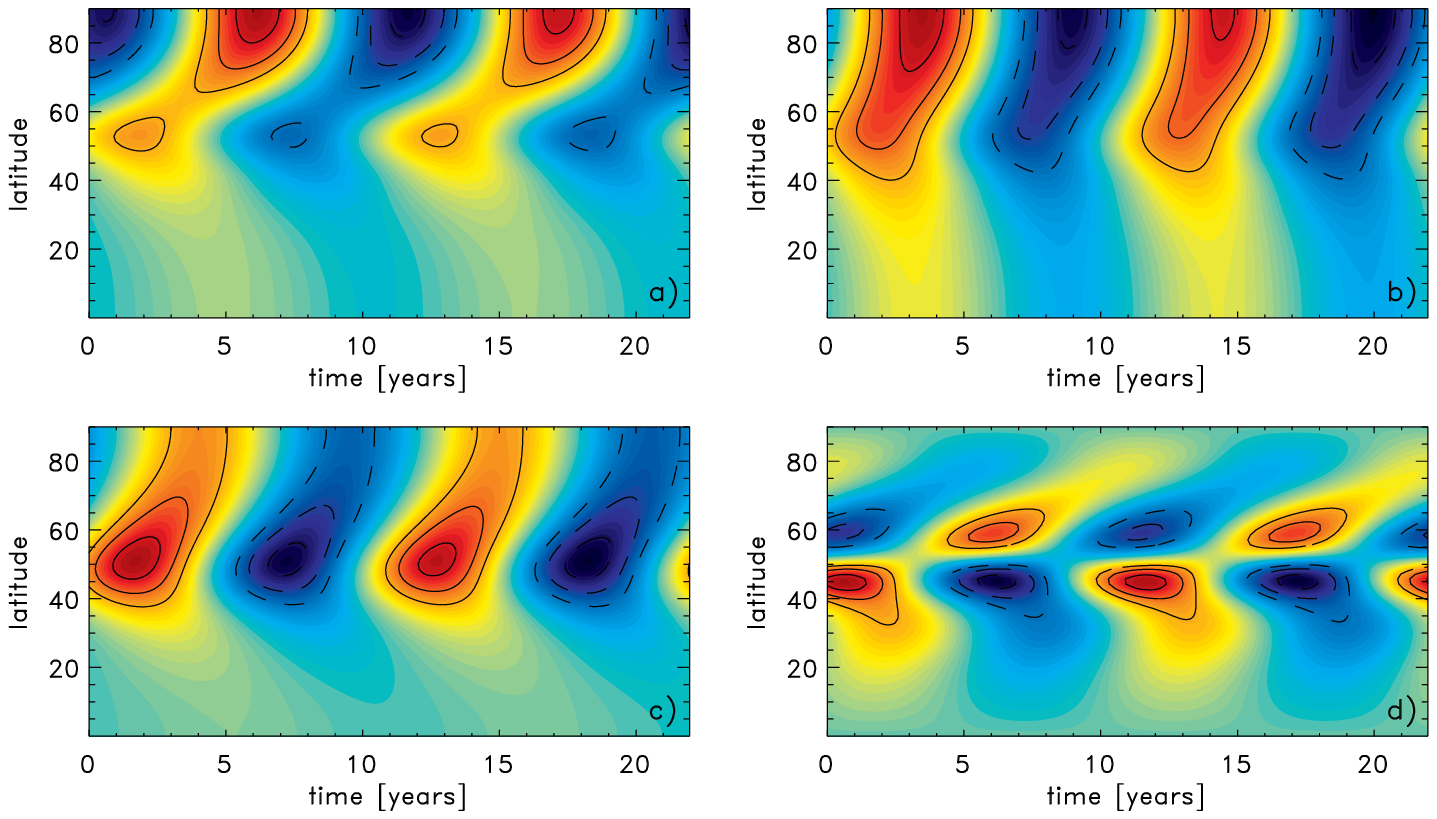}}
  \caption{Influence of viscosity and meridional flow on the poleward 
    propagation of zonal flow variations. Panels a) and b) compare to models
    different values of the turbulent viscosity: a) 
    $\nu_t=3\times 10^8\,\dunit$ and b) $\nu_t=1.2\times 10^9\,\dunit$. A higher 
    value of the turbulent viscosity leads to a faster poleward propagation of 
    the flow pattern. Both cases show the response to a periodic mechanical 
    forcing at $45\degr$ latitude. Panel c) shows the same model as panel a),
    however, the meridional is not allowed to change in response to the zonal
    flow variation. In this case the maximum amplitude coincides with the 
    location of the forcing region and the velocity of the poleward 
    propagation of the signal is reduced. Panel d) shows the variation of 
    meridional flow pattern (the reference state meridional flow is subtracted)
    associated with poleward moving zonal flow pattern shown in Panel a).
    The amplitude of the meridional flow variation is around $0.2\,\vunit$ for 
    a solar like zonal flow variation of around $4$ nHz at the pole. 
  }
  \label{f3}
\end{figure*}

Our theoretical analysis is based on the differential rotation model developed
by \citet{Rempel:2005}. This model is an axisymmetric mean field model 
incorporating a parametrization of turbulent angular momentum transport 
'$\Lambda$-effect' \citep{Kitchatinov:Ruediger:1993}. We refer to 
\citet{Rempel:2005} for details
of this model. For this investigation we use a reference model close to the
model used in \citet{Rempel:2006:dynamo}, with the minor difference of using
a larger value of $\delta=-3\times 10^{-5}$ for the superadiabaticity in the 
overshoot region (leading to more solar like differential rotation in terms
of inclination of the isorotation contours). Figure \ref{f1} shows the 
differential rotation and the meridional flow stream function for the 
reference model.

Torsional oscillations can be driven in general through mechanical forcing
(Lorentz force) or thermal forcing (pressure imbalances drive geostrophic
flows in a rotating system). Most investigations so far focused on the role of
the longitudinal Lorentz force, neglecting azimuthal components. This can be
partially justified by the fact that the toroidal field is significantly stronger
than the poloidal field in an $\alpha\Omega$ dynamo, leading also to a dominant
component of the Lorentz force in the longitudinal direction. The azimuthal component
could be important for explaining (observed) deviations from the Taylor-Proudman state
through a magnetostrophic balance; however, as we show later, this is unlikely to happen
in the bulk of the convection zone. Thermal forcing is very efficient in
that reproducing the observed amplitude requires only temperature fluctuations
of a few tenth of a degree, which can be derived from a geostrophic balance.
Assuming a latitudinal balance between Coriolis force and pressure force gives:
\begin{equation}
\Omega_0\Omega_1 r^2 \sin(2\theta)=\frac{1}{\rh}\frac{\partial p}
{\partial \theta}\approx\frac{1}{\rh}\frac{\Delta p}{\Delta \theta}
\approx\frac{p}{\rh T}\frac{\Delta T}{\Delta \theta}\;,
\label{geostrophic}
\end{equation}
which leads to
\begin{equation}
  \Delta T\approx \Delta\theta \left(\Omega_0 r\right)^2\sin(2\theta)
  \left(\frac{\Omega_1}{\Omega_0}\right)\left(\frac{R}{\mu}\right)^{-1}\;.
\end{equation}
Here $\Omega_0$ denotes the reference state rotation rate, $\Omega_1$ the
perturbation (torsional oscillation), $R$ is the gas constant, $\mu$
the mean molecular weight, $r$ the radius, and $\theta$ the co-latitude. 
Adopting solar values of 
$\Omega_0=2.7\times 10^{-6}\,\mbox{s}^{-1}$, $r=7\times 10^8\,\mbox{m}$,
$\mu=0.62$, $\Delta\theta=10\degr$ (width of active region belt) and 
$\Omega_1/\Omega_0\sim 0.005$, we get
\begin{equation}
  \Delta T\sim 0.25\,\mbox{K}\;.
\end{equation}
This value is consistent with the temperature perturbation of $0.1$ to $0.2$ K
\citet{Rempel:2006:dynamo} imposed in their model to drive the low latitude
branch of torsional oscillations.
We emphasize that the total temperature perturbation is nearly independent
of depth for a fixed torsional oscillation amplitude, meaning that the relative
perturbation close to the surface is around $5\times 10^{-5}$ 
(the same value was given by \citet{Spruit:2003}), but only around $10^{-7}$
close to the base of the convection zone.

The following questions will be addressed in this investigation:
\begin{enumerate}
  \item[$\bullet$] Are torsional oscillations of thermal or mechanical origin? 
  \item[$\bullet$] Are they a consequence of traveling or periodic 
    perturbations in the solar convection zone?
\end{enumerate}

We focus our discussion separately on the high latitude and low latitude 
branch for reasons that will become more evident in the following discussion.

\section{High latitude branch}
The high latitude branch has a larger amplitude than the low latitude branch
and is propagating poleward, while the low latitude branch is clearly following
the equatorward propagating magnetic activity belt. This leads to the 
fundamental question of whether the poleward branch is also associated with
a poleward propagating magnetic pattern at the base of the convection zone
(as it has been seen in many interface dynamos), which is not visible at the
solar surface. We cannot rule out such a possibility here, but we will show
that this is not conclusive and the poleward propagation can be explained
as a response of the coupled differential rotation meridional flow system to
a non-traveling periodic perturbation in mid-latitudes. To demonstrate this,
we incorporate in the differential rotation model a mechanical (in the equation
for $\Omega$) or thermal (in the entropy equation) forcing term of the form
\begin{eqnarray}
\frac{\partial \Omega_1}{\partial t}&=&\ldots+A_{\Omega} f(r,\theta)
\sin(\omega_c t)\\
\frac{\partial s_1}{\partial t}&=&\ldots+A_{s} f(r,\theta)
\sin(\omega_c t)\;.
\end{eqnarray}    
Here $\omega_c$ corresponds to a $11$ year periodicity (as observed for 
torsional oscillations), $A_{\Omega}$ and $A_{s}$ are the amplitudes of the
forcing, and $f(r,\theta)$ is a Gaussian profile of the form:
\begin{equation}
  f(r,\theta)=\exp\left[-\left(\frac{r-r_0}{\Delta r}\right)^2\right]\cdot
    \exp\left[-\left(\frac{\theta-\theta_0}{\Delta \theta}\right)^2\right]\;.
\end{equation}
We use for all following models the parameters $r_0=0.85\,R_{\odot}$,
$\Delta r=0.05\,R_{\odot}$, $\Delta \theta=0.125$, and vary the co-latitude
between $\pi/6$, $\pi/4$, and $\pi/3$. For the results presented here the
amplitudes $A_{\Omega}$ and $A_{s}$ are of secondary concern (as long as
$\Omega_1\ll \Omega_0$). We have chosen values leading to torsional 
oscillations of a few percent of the rotation rate.

\begin{figure*}
  \resizebox{\hsize}{!}{\includegraphics{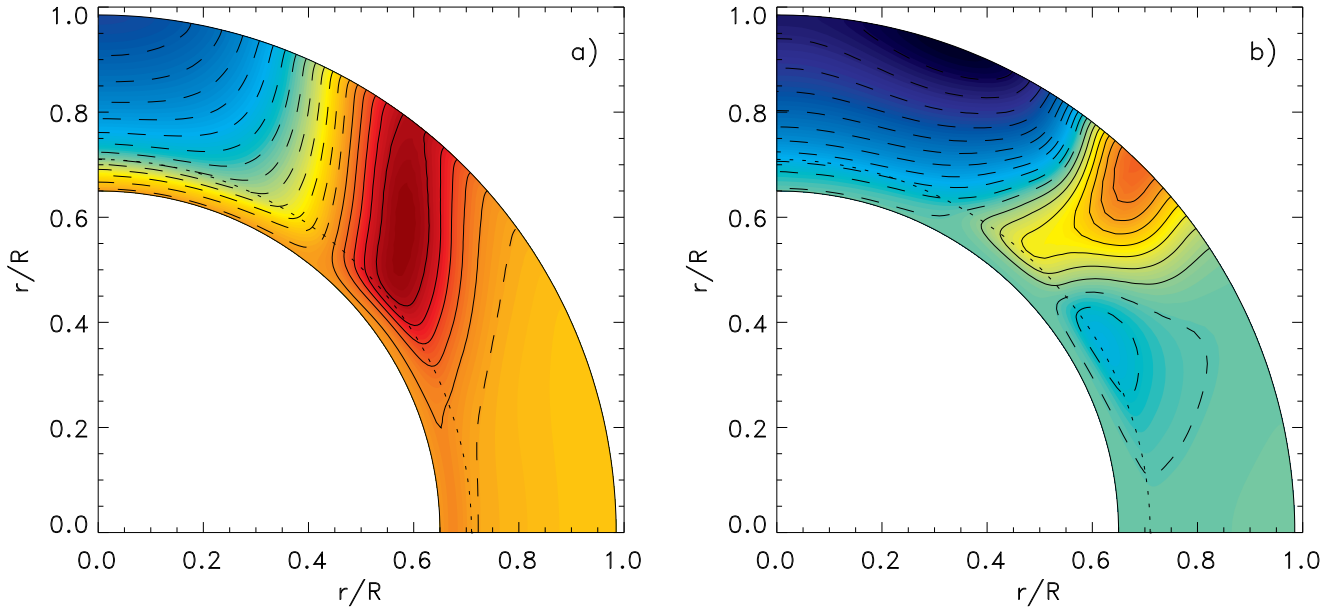}}
  \caption{Snapshots of the torsional oscillations caused by mechanical
    forcing (left) and thermal forcing (right). The snapshots correspond 
    to $t=0$ years in Fig. \ref{f2}.
    Torsional oscillations caused by thermal forcing have the tendency
    to show stronger deviations from the Taylor-Proudman state.
  }
  \label{f4}
\end{figure*}

Figure \ref{f2} shows on the left side models with mechanical forcing and on 
the right side models with thermal forcing. From top to bottom, the latitude 
of the forcing is varied between $60\degr$, $45\degr$, and $30\degr$.
Independent of the nature of the periodic forcing, all models show a strong
poleward propagating pattern and a very weak equatorward propagating pattern 
away from the latitude at which the forcing is applied. The equatorward pattern
increases in amplitude when the location of the forcing is moved closer to
the equator. The pattern typically shows two peaks, one close to the latitude
of the forcing and one right at the pole (with larger amplitude). Comparing
the latitudinal extent of the pattern on the polar side of the forcing region
with observations, a forcing location in mid-latitudes seems most reasonable.

The time for the signal to travel from about $60\degr$ to the pole is
$3$ to $4$ years, which is close to the observed propagation speed. It turns 
out that the travel time is primarily affected by the turbulent viscosity 
assumed in the reference model, which is shown in Figure \ref{f3}. To this end
we compare two models with the viscosity values of $3\times 10^8\,\dunit$ 
(panel a) and $1.2\times 10^9\,\dunit$ (panel b). We have adjusted in each 
model the amplitude and the direction of the turbulent angular momentum 
transport ($\Lambda$-effect) such that the differential rotation and 
meridional flow remain roughly unchanged. The propagation time for the signal 
drops from around $3$ to $1$ year when the turbulent viscosity is increased
by a factor of $4$.  Since solar observations show a time delay of $3$ 
to $4$ years, this sets constraints on the value of the turbulent viscosity.
In our model we get the best fit for values $\lesssim 3\times 10^8\,\dunit$.
This value is around one order of magnitude smaller than mixing-length 
estimates for the solar convection zone, it agrees however with the results
of \citet{Kitchatinov:etal:1994} taking into account rotational quenching
of turbulent viscosity. 
Panel c) shows the same model as panel a) but the meridional flow is
not allowed to change in response to the zonal flow variation (we solve only 
the equation for $\Omega_1$ and keep the meridional flow fixed).
This leads to a significantly different flow pattern: the 
maximum amplitude is found close to the forcing region and the propagation of
the signal toward the pole slows down. In this way the variation of the
meridional flow has a significant influence on the structure of the torsional
oscillation pattern. Interestingly, the meridional flow of the reference state
is not important, only the induced flow variation. Constructing a model
with the same value of the turbulent viscosity, but a different meridional 
flow, leads to a nearly identical zonal flow pattern (difference in the $1\%$ 
range). Panel d) shows the meridional flow variation associated with the 
poleward moving zonal flow pattern shown in panel a). A solar like amplitude 
of around $4$ nHz for the torsional oscillation leads to a meridional flow 
variation of around $0.2\,\vunit$.

To our knowledge this is so far the only way to constrain the turbulent 
viscosity directly by observations, without having to relate it to the
magnetic diffusivity by making assumptions about the magnetic Prandtl number. 
We emphasize that our model
assumes a constant value throughout the convection zone. In models with a depth
dependent viscosity the above mentioned value should reflect more the values 
in the lower half of the convection zone rather than the surface layers, 
since they contribute more to the angular momentum transport due to their 
larger density.

From the results shown in Figure \ref{f2} it is difficult to judge whether it
is possible to distinguish between a mechanical or thermal forcing. Figure
\ref{f4} shows meridional cross-sections of the zonal flow pattern. The most
obvious difference here is that the mechanical forcing leads to torsional
oscillations more aligned with the rotation axis (Taylor-Proudman theorem),
while thermal forcing shows a tendency to produce more radially aligned 
patterns. 
Beside aligning structures with the rotation axis, the Taylor-Proudman 
constraint also leads to an almost constant amplitude within the convection 
zone, while in the case of thermal forcing the entropy perturbations allow
for more variation of the amplitude as function of depth.

In the case of the high latitude branch the observations are not detailed
enough to distinguish between thermal and mechanical forcing (this may be 
because in high latitudes the radial direction almost coincides with the 
axis of rotation).
We will show in the next section that there is a significant difference 
between mechanical and thermal forcing effects in low latitudes.

\section{Low latitude branch}

\begin{figure*}
  \resizebox{\hsize}{!}{\includegraphics{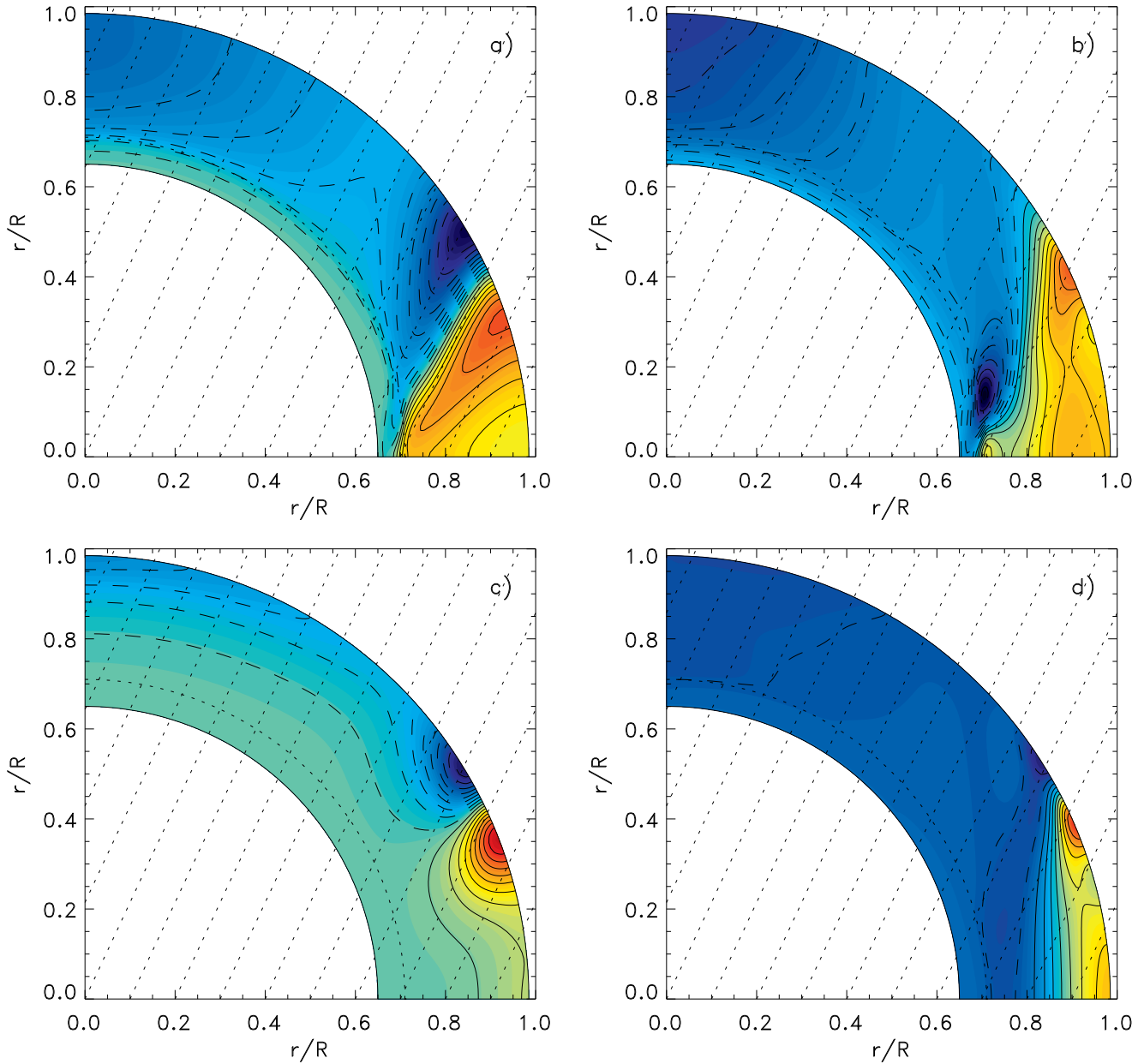}}
  \caption{Mechanical forcing of low latitude zonal flow variations. Panels
    a) and c) show the forced zonal flow considering no change of the
    meridional flow (solving the equation for $\Omega_1$ for the fixed
    reference state flow), while panels b) and d) show the zonal flow
    solving the full system.  Panels a) and b) use a forcing function with
    $r_0=0.75\,R_{\odot}, \theta_0=80\degr$, c) and d) with
    $r_0=0.95\,R_{\odot}, \theta_0=65\degr$.
    While panels a) and c) show a zonal flow that closely resembles the
    properties of the forcing function [especially the $25\degr$ inclination
    in panel a)], the results shown in panels b) and d) are strongly influenced
    by rotation. The flow pattern becomes strongly aligned with the axis
    of rotation and also the amplitude of the zonal flow band with slower
    rotation is significantly smaller than the faster zonal band.
  }
  \label{f5}
\end{figure*}

\begin{figure*}
  \resizebox{\hsize}{!}{\includegraphics{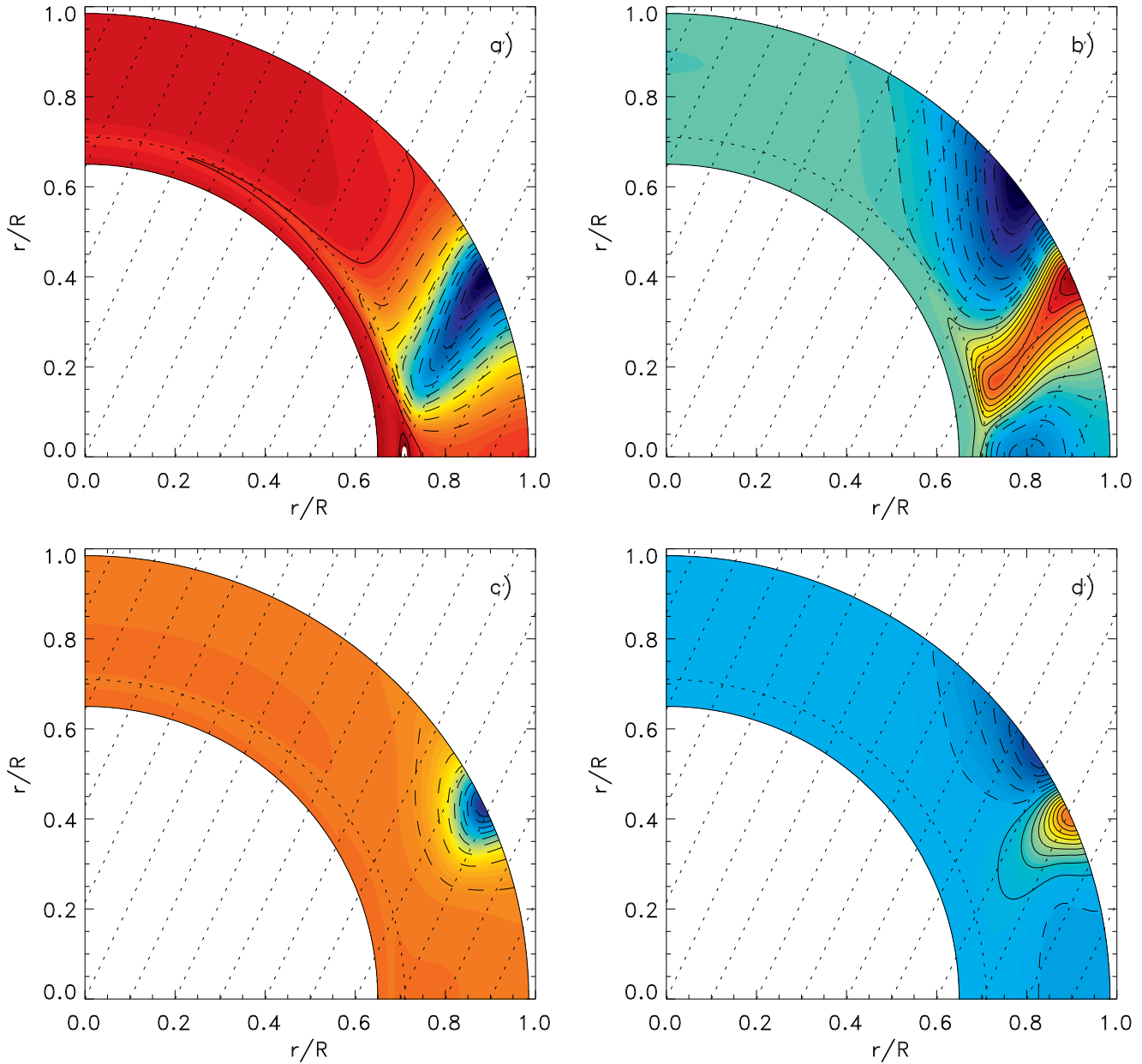}}
  \caption{Thermal forcing of low latitude zonal flow variations. The panels
    a) and c) show the imposed entropy perturbation, the panels b) and d)
    the resulting zonal flows. Panels a) and b) use a forcing function with
    $r_0=0.75\,R_{\odot}, \theta_0=80\degr$, c) and d) with
    $r_0=0.95\,R_{\odot}, \theta_0=65\degr$.
    Contrary to the results presented in Figure \ref{f5} for mechanical 
    forcing, the driven zonal flows show significant deviations from the
    Taylor-Proudman state. Especially in panel b) the imposed $25\degr$ angle
    of the thermal forcing function is clearly visible in the zonal flow field.
    Also both zonal bands (faster and slower one) are visible, with the faster
    band on the equatorial side being of larger amplitude and more confined.
  }
  \label{f6}
\end{figure*}

\begin{figure}
  \resizebox{8.25cm}{!}{\includegraphics{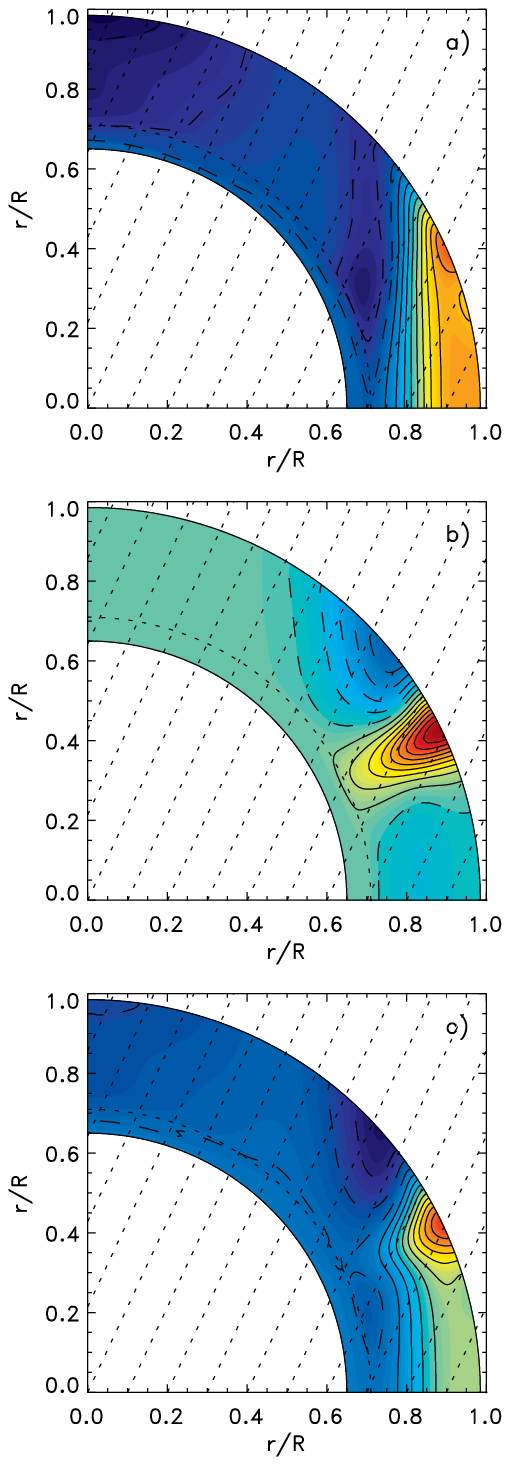}}
  \caption{Combination of mechanical and thermal forcing: a) pure mechanical
    forcing, b) pure thermal forcing, c) combination of both. The mechanical 
    and thermal forcing functions used here have a radial alignment and the
    amplitude of the zonal flow in a) and b) is roughly the same.
    The combination of both forcings produces roughly a $25\degr$ inclination 
    as observed in the solar case. 
  }
  \label{f7}
\end{figure}

The most striking feature of the low latitude branch is the systematic
deviation from the Taylor-Proudman state. As first pointed out by
\citet{Howe:etal:2004:phase}, the lines of constant phase show an inclination
of $25\degr$ to the axis of rotation, similar to the isorotation contours
of the differential rotation. We present here a few idealized experiments to
differentiate between possible mechanical and thermal forcing. To this end
we construct a mechanical forcing function and a cooling function that has a
$25\degr$ inclination angle with respect to the axis of rotation. For reasons of
simplicity we apply here a stationary perturbation for a time interval of
a few month to illustrate the effect. The forcing functions are given by:
\begin{eqnarray}
\frac{\partial \Omega_1}{\partial t}&=&\ldots+A_{\Omega}\, g(r,\theta)
\left[\theta-\zeta(r)\right]\label{mech_lowlat}\\
\frac{\partial s_1}{\partial t}&=&\ldots+A_{s}\, g(r,\theta)
\label{therm_lowlat}\;,
\end{eqnarray}    
with
\begin{eqnarray}
  g(r,\theta)&=&0.5\left[1+\tanh\left(\frac{r-r_0}{\Delta r}\right)\right]
  \nonumber\\
  &&\cdot\exp\left[-\left(\frac{\theta-\zeta(r)}{\Delta \theta}\right)^2\right]
  \,.
\end{eqnarray}
The function $\zeta(r)$ specifies the inclination of the forcing with respect 
to the rotation axis. Assuming that $\theta_0$ is the location of the forcing
at $r_0$ and $\lambda$ is the inclination angle we get:
\begin{equation}
  \zeta(r)=\lambda+\arcsin\left[\frac{r_0}{r}\sin(\theta_0-\lambda)
    \right]\label{phase}\;.
\end{equation}
Here $\theta_0$ determines the co-latitude of the perturbation at the radius
$r_0$, where $r_0$ marks the depth at which the forcing decreases to half of
the convection zone value. We use in the following discussion common values 
of $\Delta r=0.05\,R_{\odot}$, $\lambda=25\degr$, and $\Delta\theta=0.1$
($\Delta\theta=0.05$) for mechanical (thermal) forcing, respectively.

To compare results for forcing functions extending through the entire 
convection zone with those from surface forcing functions, we will use in 
both cases the parameters
$r_0=0.75\,R_{\odot}, \theta_0=80\degr$ (co-latitude); and $r_0=0.95\,R_{\odot}, 
\theta_0=65\degr$. The amplitudes $A_{\Omega}$ and $A_{s}$ are chosen so
that applying the forcing for a time interval of about $4$ months leads to
an amplitude of the zonal flow of a few nHz. All results shown in the 
following discussion show the zonal flow after $4$ months of mechanical or
thermal forcing.

The forcing function for the
mechanical forcing Eq. (\ref{mech_lowlat}) is multiplied by the term 
$\left[\theta-\zeta(r)\right]$ in order to produce a perturbation that has a 
faster zonal flow equatorward and slower zonal flow poleward of the region
the forcing is applied. In the case of the thermal forcing this is not 
required, since the response to a cooling automatically leads to the formation
of two geostrophic flows with opposite directions.

Figure \ref{f5} summarizes the results obtained with mechanical forcing,
Figure \ref{f6} the results from thermal forcing. In the case of mechanical
forcing the zonal flows do not resemble the properties of the forcing function
(especially the $25\degr$ inclination) unless meridional flow variations are
completely suppressed as shown in Figure \ref{f5}a) and c). If meridional 
flow variations
are considered, the zonal flow pattern shows strong influence of rotation,
aligning the pattern with the axis of rotation (Taylor-Proudman theorem).
As a consequence, the resulting flow is entirely dominated by the influence
of rotation and does not resemble the properties of the applied forcing
function. Since the pattern has the tendency to spread parallel to the 
axis of rotation, a perturbation traveling at the base of the convection zone
($r=0.72\,R_{\odot}$) from $30\degr$ latitude toward the equator, would 
produce close to the surface a pattern moving from around $40\degr$ to 
$20\degr$ and could therefore not explain the observed low latitude branch of 
the observed torsional oscillations. Even applying the forcing close to the
surface does not lead to a reasonable flow pattern as shown in Figure 
\ref{f5}d).

This result is not too surprising, since adding a forcing in the $\phi$
direction does not impact the meridional force balance that leads to the
Taylor-Proudman state. This raises the question if additional magnetic 
stresses in the meridional plane could cause a deviation from the 
Taylor-Proudman state as observed (magnetostrophic balance). 
In a typical $\alpha\Omega$-model for the solar dynamo, the poloidal 
field is at least a factor of $100$ weaker than the toroidal field. 
Having a toroidal field strength of around $1$ T ($10$ kG) 
\citep{Rempel:2006:dynamo}, this yields around 
$0.01$ T ($100$ G) for the poloidal field. An estimate similar to 
Eq. (\ref{geostrophic}) leads in the case of a magnetostrophic balance to
\begin{equation}
  B_p^2\approx \mu_0\rh \Delta\theta \left(\Omega_0 r\right)^2\sin(2\theta)
  \left(\frac{\Omega_1}{\Omega_0}\right)\;.
\end{equation}
Here we assumed that the $\theta$-component of the Lorentz force can be 
estimated as $B_p^2/(\mu_0\,r\,\Delta\theta)$, where $B_p$ denotes the 
poloidal field strength.
Adopting solar values of $\Omega_0=2.7\times 10^{-6}\,\mbox{s}^{-1}$, 
$r=7\times 10^8\,\mbox{m}$, $\Delta\theta=10\degr$, 
$\Omega_1/\Omega_0\sim 0.005$, and we get
\begin{equation}
  B_p\approx 1\,\mbox{T}\sqrt{\frac{\rh}{\rh_{\rm bc}}}\;,
\end{equation}
where $\rh_{\rm bc}=200\,\mbox{kg}\,\mbox{m}^{-3}$ denotes the density at the 
base of
the convection zone. These values indicate that a magnetostrophic
balance is very unlikely in the bulk of the convection zone, but very close to
the surface, where the density is very low, it cannot be ruled out (for a 
typical solar model we have $\rh\approx 10^{-2}\rh_{\rm bc}$ at 
$r=0.98\,R_{\odot}$ and $\rh\approx 10^{-4}\rh_{\rm bc}$ at 
$r=0.997\,R_{\odot}$). There are additional terms of the Lorentz force in the
meridional plane that include the much stronger toroidal field: the gradient of
the magnetic pressure and magnetic tension force arising from spherical 
geometry $\sim B_{\phi}^2/r$. The latter can lead to a prograde jet within
the magnetized region as discussed by \citet{Rempel:etal:2000} and 
\citet{Rempel:Dikpati:2003}, but not to a 
flow pattern with two opposite flows on both sides of the active region belt.

In order to address the influence of magnetic pressure, we show in Eq. 
(\ref{tpt}) the $\phi$ component of the vorticity equation considering 
a stationary flow and neglecting viscous stresses (we also neglect here the
magnetic tension force for the reason mentioned above):
\begin{equation}
  r\sin\theta\frac{\partial \Omega^2}{\partial z}
  =\frac{g}{\gamma r}\frac{\partial}{\partial \theta}
  \left(s_1+\frac{p_m}{p_0}\right)
  \label{tpt}
\end{equation}   
Note that this equation addresses the deviation of the full differential 
rotation from Taylor-Proudman state. 
Therefore $s_1$ also contains entropy perturbations associated 
with the reference state differential rotation. Writing $\Omega=\Omega_r+\Omega_t$
and $s_1=s_r+s_t$, where the subscript 'r' refers to the reference state and 
the subscript 't' to the torsional oscillation part, we get in
leading order:
\begin{equation}
  2\,r\sin\theta\left(\Omega_r\frac{\partial \Omega_t}{\partial z}+
  \Omega_t\frac{\partial \Omega_r}{\partial z}\right)
  =\frac{g}{\gamma r}\frac{\partial}{\partial \theta}
  \left(s_t+\frac{p_m}{p_0}\right)\label{tpt2}
\end{equation} 
Here the first term on the left hand side is the dominant one (typically at
least a factor of $5$ larger than the second one for the examples discussed 
here), relating directly the $z$ derivative of $\Omega_t$ to the entropy 
perturbation and magnetic pressure. Therefore, if the right hand side of 
Eq. (\ref{tpt2}) is zero, the torsional oscillation pattern has to be 
very close to the Taylor-Proudman state with $\partial\Omega_t/\partial z=0$. 
The occurrence of $s_t$ together with $p_m/p_0$ means that a toroidal magnetic 
field with neutral magnetic buoyancy does not lead to any deviation from the 
Taylor-Proudman state. If magnetic buoyancy is not compensated, 
magnetic pressure causes a flow 
pattern with two opposite flows on both sides of the active region belt; 
however, they would have the wrong sign (fast moving band on poleward side, 
slow moving on equatorward side; see also the discussion in next paragraph) 
when we make the reasonable assumption that the active region belt 
has a higher magnetic pressure than the surrounding region. We will show in 
the following paragraph that a reasonable flow perturbation results from the 
assumption that the active region belt has lower gas pressure, which can be the
consequence of a thermal perturbation.
 
The zonal flow presented in Figure \ref{f6}b) shows
the $25\degr$ inclination angle imposed through the thermal forcing function.
Even though the forcing function is symmetric with respect to the 
latitude of the maximum (given by $\theta=\zeta(r)$), the resulting zonal
flows show a significant asymmetry, with the equatorward faster rotating band
more confined than the poleward slower rotating band. This is a consequence of
the fact that Eq. (\ref{tpt}) considers the radial and the latitudinal 
component of the Coriolis force together. The relative contribution of both 
components depends on the latitude. In high latitudes we have 
$\partial/\partial z\sim \partial/\partial r$, resulting in a symmetric flow 
pattern with the slower rotating band poleward and the faster rotating band
equatorward, while in low latitudes we have 
$\partial/\partial z\sim -\partial/\partial \theta$ leading to a solution
$\Omega_t\sim -s_t$, showing only a fast rotating band centered around the
cooling region. In mid-latitudes the solution is a combination of both, with
having a stronger fast rotating band that is also more close to the center of 
the cooling region. Applying the thermal 
forcing only to the surface, as shown Figure \ref{f6}d), produces a zonal 
flow perturbation that is much more confined to the surface; the $25\degr$ 
inclination is less visible in this case. 

This study shows that it is much easier to produce a zonal flow pattern with
the observed properties of the low latitude torsional oscillations through
thermal forcing rather than mechanical forcing, but it does not explain the
observed $25\degr$ inclination angle, since this angle has to be imposed on
the thermal forcing function. This could be a consequence of anisotropic 
thermal heat diffusivity that is larger parallel than perpendicular to the 
axis of rotation \citep{Kitchatinov:etal:1994}. A thermal signal imposed at 
the surface, as suggested by \citet{Spruit:2003}, would therefore not 
penetrate radially into the convection zone but rather in an angle more
aligned with the rotation axis. Another possible explanation could be
a combination of mechanical and thermal forcing. To illustrate this, we
computed in Figure \ref{f7} a zonal flow resulting from both types
of forcing. Panel a) shows the result of the mechanical forcing alone,
panel b) the result of the thermal forcing. In both cases we used
$r_0=0.9\,R_{\odot}$ and $\theta_0=25\degr$. The forcing functions show a
radial alignment ($\zeta(r)=\theta_0={\rm const.}$). The amplitude of the
zonal flow in a) and b) is roughly the same. Applying mechanical and thermal
forcing together results in a zonal flow with roughly $25\degr$ inclination
with respect to the axis of rotation.

In this section we focused entirely on static forcing functions, not 
addressing the equatorward propagation of the signal in the course of the
solar cycle. This could be easily addressed by incorporating a time dependent
phase into Eq. (\ref{phase}); however, this does not impact the conclusions of
this section concerning the importance of the Taylor-Proudman theorem, since
the timescale for the establishment of the Taylor-Proudman state is 
significantly shorter than the solar cycle.

\section{Discussion}
We present in this paper a series of numerical experiments addressing the
question of what type of forcing is required to produce a zonal flow variation
showing the properties of the observed solar torsional oscillations. Since this
study is independent from a specific dynamo model, we can infer a few general
conclusions about torsional oscillations and their possible origin:
\begin{enumerate}
\item[$\bullet$] The Taylor-Proudman theorem applies also to perturbations of
  $\Omega$. Therefore it is crucial to solve in models addressing the origin
  of torsional oscillations the full momentum equation, including variations
  of the meridional flow.
\item[$\bullet$] Mechanically forced zonal flows always have the tendency to 
  spread through the entire convection zone parallel to the axis of rotation. 
\item[$\bullet$] Significant deviations from the Taylor-Proudman state (as 
  observed in low latitudes) require thermal perturbations. A pure mechanical
  origin of the low latitude torsional oscillations is very unlikely.
\item[$\bullet$] The fact that the phase of the torsional oscillation signal 
  in low latitudes is inclined such that the signal occurs for a fixed latitude
  at the base of the convection around 2 years prior to the surface signal
  does not indicate necessarily that the origin has to be at the base of the 
  convection zone. This is a consequence of the influence of rotation.
\item[$\bullet$] The high latitude branch of torsional oscillations does not
  require a forcing propagating poleward in the course of the solar cycle.
  A poleward propagating zonal flow pattern is the natural response of the
  coupled differential rotation / meridional flow system to a periodic
  forcing (mechanical or thermal) in mid-latitudes. 
\item[$\bullet$] The poleward propagation speed of the torsional oscillation
  pattern is only
  determined by the value of the turbulent viscosity of the reference model.
  Best agreement is found for values of $\nu_t$ less than $3\times 10^8\,\dunit$.
\end{enumerate}

In this paper we did not consider the physical origin of the imposed 
mechanical or thermal forcings. Possible physical explanations for the
forcings considered in this paper are the following:

\vspace{0.25cm}
\noindent
{\it Macroscopic Lorentz force:}

Recently \citet{Rempel:2006:dynamo} showed that the macroscopic Lorentz force
in a flux transport dynamo leads to a periodic source in mid-latitudes, where
the poloidal magnetic field is sheared by the differential rotation while
getting transported downward by the meridional flow. The model provides a
poleward propagating branch with the correct amplitude and phase relation to
the magnetic cycle. Unlike some of the processes listed below, this type of
feedback is unavoidable for a dynamo that is energetically driven 
through differential rotation. \citet{Rempel:2006:dynamo} showed that the 
amplitude of the solar torsional oscillations is consistent with a dynamo that
converts around $0.1\%\,L_{\odot}$ of energy and produces around $15$ kG of
toroidal field at the base of the convection zone.

A different model was presented recently by 
\citet{Covas:etal:2000,Covas:etal:2004,Covas:etal:2005}. Their model
uses the observed solar rotation profile and allows for perturbations around 
this profile by considering the longitudinal component of the macroscopic 
Lorentz force, while the azimuthal component of the momentum equation is neglected 
(no meridional flow is considered). Their results show
torsional oscillations with both branches driven by the Lorentz force that
have a remarkable agreement with observations \citep{Vorontsov:etal:2003}.
Even though most of the magnetic activity is in their model concentrated in
low latitudes, there are also magnetic field patterns moving poleward 
together with the torsional oscillation pattern. We showed in this paper
that a poleward propagating magnetic pattern is not required to drive the
high latitude torsional oscillations, but on the other hand we cannot rule
out such a possibility based on the results we presented here. It would
require a more detailed analysis of their model to determine whether
the poleward propagating magnetic activity is the main driver or a response
to a periodic forcing in mid-latitudes as shown by \citet{Rempel:2006:dynamo}.

As far as it concerns the low latitude branch of torsional oscillations 
we expect a significant influence of rotation, which is not present in a 
model not considering self-consistently the meridional flow. The alignment
of the phase with the rotation axis makes it difficult to obtain a surface 
pattern of torsional oscillations propagating all the way to the equator 
especially if the Lorentz force action takes place in the lower part of the 
convection zone (which is the case in a dynamo model living mainly on the 
radial shear in the tachocline). This situation is comparable to the development
of mean field models for differential rotation more than a decade ago. While early 
models that were not considering the meridional flow got a remarkable agreement with 
observations, later models solving all components of the momentum equation ended up 
in the Taylor-Proudman state. The problem is still under discussion, even though most
people agree now that thermal perturbations explain the observed profile of the solar
differential rotation \citep{Kitchatinov:Ruediger:1995,Rempel:2005,Miesch:etal:2006}.
In our experience it is much more difficult to explain the low latitude torsional
oscillations when the azimuthal components of the momentum equation are considered.
The situation is different for the high latitude branch, since there the alignment
of perturbations with the axis of rotation does not impose such a strong constraint.

\vspace{0.25cm}
\noindent
{\it Enhanced radiative loss in active region belt:}

\cite{Spruit:2003} proposed that the torsional oscillations are a response to
enhanced radiative losses in the active region belt due to small scale magnetic
flux elements. \citet{Rempel:2006:dynamo} parametrized this idea and showed 
that the cooling of the active region belt drives a zonal flow that is, at 
least close to the surface, in good agreement with observations.
Also meridional flow changes, corresponding to an inflow into the active 
region belt, reasonably agree with observations \citep{Komm:etal:1993,
Komm:1994,Zhao:Kosovichev:2004}. However, the flow pattern deeper in
the convection zone does not show the correct phase relation and depth 
dependence, which is most likely a consequence of the simple diffusive 
treatment of thermal perturbations. Better agreement could be achieved if
the thermal perturbation imposed in the surface layers penetrates deep enough
and is also influenced by anisotropic diffusivity to yield the observed
inclinations with respect to the axis of rotation. 

\vspace{0.25cm}
\noindent
{\it Quenching of convective energy flux:}

Due to the very small amplitude of temperature perturbations required for 
driving zonal flows it seems conceivable that quenching of convective energy 
flux by the dynamo generated field can also cause thermal shadows within the
convection zone resulting in zonal flows. Contrary to the process proposed by
\citet{Spruit:2003}, these temperature perturbations would originate close to 
the base of the convection zone, where the strongest magnetic field is found.
Again, an anisotropic convective energy flux would be required to explain the
observed phase relation.

\vspace{0.25cm}
\noindent
{\it Quenching of turbulent angular momentum flux ($\Lambda$-quenching):}

Microscopic Lorentz force feedback in terms of quenching of turbulent 
transport processes driving differential rotation has been addressed by
\citet{Kitchatinov:Pipin:1998}, \citet{Kitchatinov:etal:1999}, and
\citet{Kueker:etal:1999} as
possible explanation for torsional oscillations. Since $\Lambda$-quenching
leads to an additional source term in the equation for $\Omega$, this is
a mechanical forcing similar to macroscopic Lorentz force feedback. Therefore
all the results discussed above for mechanical forcing also apply in this
case.

\vspace{0.25cm}
It has been speculated whether the similar inclination of the phase of the
torsional oscillation pattern and the isorotation contours is a coincidence
or not. None of the processes discussed here would necessarily lead to a 
similar angle; however, the physical cause behind both, a thermally induced
deviation from the Taylor-Proudman state, is similar.

\acknowledgements
This work originated from discussions with Manfred Sch\"ussler during a visit
of the author at the Max-Planck Institute for Solar-System-Research in 
Katlenburg-Lindau, Germany. Very helpful comments by M. Sch\"ussler, 
K.~B. MacGregor, P.~A. Gilman and the anonymous referee are gratefully 
acknowledged.

%\bibliographystyle{natbib/apj}
%\bibliography{natbib/apj-jour,natbib/papref}

%\bibliography{ms}

\end{document}